\documentclass[12pt]{iopart}
\usepackage{iopams}  
\usepackage{graphicx}
\newcommand{\lf}{\left}
\newcommand{\rg}{\right}
\newcommand{\be}{\begin{equation}}
\newcommand{\ee}{\end{equation}}
\newcommand{\bea}{\begin{eqnarray}}
\newcommand{\eea}{\end{eqnarray}}
\newcommand{\ba}{\begin{array}}
\newcommand{\ea}{\end{array}}
\newcommand{\bd}{\begin{displaymath}}
\newcommand{\ed}{\end{displaymath}}

\newcommand{\g}{\gamma}
\newcommand{\G}{\Gamma}
\newcommand{\D}{\Delta}
\renewcommand{\d}{\delta}

\newcommand{\s}{\sigma}
\newcommand{\vf}{\varphi}
\newcommand{\ve}{\varepsilon}

\renewcommand{\o}{\omega}

\renewcommand{\k}{\kappa}
\newcommand{\nn}{\nonumber}
\newcommand{\ra}{\rangle}
\newcommand{\la}{\langle}
\newcommand{\hf}{\frac{1}{2}}

\begin{document}

\title[Lasing in Circuit QED]{Few-Qubit Lasing in circuit QED }
\author{Stephan Andr\'e$^{1,4}$, Valentina Brosco$^{2}$, Michael Marthaler$^{1,4}$, Alexander 
Shnirman$^{3,4}$, Gerd Sch\"on$^{1,4}$}

\address {$^1$Institut f\"ur Theoretische Festk\"orperphysik,
       Karlsruhe Institute of Technology, 76128 Karlsruhe, Germany}
\address{$^2$Dipartimento di Fisica, Universit\`a ``La Sapienza'', P.le 
A. Moro 2, 00185 Roma, Italy}
\address{$^3$Institut f\"{u}r Theorie der Kondensierten Materie, 
       Karlsruhe Institute of Technology, 76128 Karlsruhe, Germany}
\address{$^4$DFG Center 
for Functional Nanostructures (CFN), Karlsruhe Institute of Technology, 76128 
Karlsruhe, Germany}       

\ead{schoen@kit.edu}
\begin{abstract}
Motivated by recent experiments, which demonstrated lasing and 
cooling of the electromagnetic modes in a resonator 
coupled to a superconducting qubit, we describe the specific 
mechanisms creating the population inversion, and we 
study the spectral properties of these systems in the lasing state.
Different levels of the theoretical description, i.e., the 
semi-classical and the semi-quantum approximation, as well as an analysis 
based on the full Liouville equation are compared. We extend the usual 
quantum optics description to account for strong qubit-resonator 
coupling and include the effects of low-frequency noise. 
Beyond the lasing transition we find for a single- or 
few-qubit system the phase diffusion strength to grow with the coupling strength, which in turn deteriorates the lasing state.  

\end{abstract}

\pacs{85.25.Cp 42.50.Pq 03.65.Yz} \maketitle

\section{Introduction}
The search for efficient coupling and read-out architectures of scalable 
solid-state quantum computing systems has opened 
a new field, called ``circuit QED" \cite{blais04}. It is the 
on-chip analogue of quantum optics ``cavity QED", 
 with superconducting qubits playing the role of (artificial) atoms
and  an electromagnetic resonator replacing the cavity.
 The resonators can be used to  
read out  the qubit state 
\cite{ilichev03,wallraff04,chiorescu,johansson06} or to couple  
qubits to perform single- and two-qubit 
gates \cite{sillanpaa,leek07,filipp,dicarlo}.

Apart from these applications for quantum information processing, 
circuit QED offers the possibility to study effects known from 
quantum optics  in electrical circuits:  
Fock states of the electromagnetic field were created and 
detected~\cite{hofheinz}, and the coherent control of photon propagation 
via electromagnetically induced transparency was shown~\cite{sillanpaa09}. 
In addition, lasing and cooling of the 
electromagnetic field in the resonator has been demonstrated: 
By creating a population inversion in a driven superconducting 
single-electron transistor 
(SSET) coupled capacitively to a microstripline resonator, 
Astafiev \emph{et al.}~\cite{astafiev} could excite a lasing peak 
in the spectrum. In another experiment,
Grajcar \emph{et al.}~\cite{grajcar}  coupled a driven flux 
qubit to a low-frequency  LC resonator and observed both 
cooling and  a tendency towards lasing via  
the so-called Sisyphus mechanism. 

In contrast to conventional lasers where many atoms are  
coupled weakly to the light field in a Fabry-Perot cavity, in the 
micromasers realized, e.g., in Refs.~\cite{astafiev,grajcar} 
a single superconducting qubit is coupled
strongly  to the microwave field in the resonator. 
Compared to conventional lasers one expects for single-atom lasers a 
lower intensity  of the radiation but stronger fluctuation
effects. Specifically, the quantum fluctuations of the 
photon number associated with spontaneous emission, which are known to 
lead to the \emph{phase diffusion} of the laser field \cite{haken}, 
have more pronounced consequences. 
As a result, even in the lasing state, phase coherence is lost after a 
 characteristic time $\tau_d$, which sets a limit 
on the linewidth of the laser radiation and thus on the visibility 
of the lasing signal. Phase diffusion was observed experimentally 
in a single-qubit maser in Ref.~\cite{astafiev}; the dependence of 
the phase diffusion on the coupling strength was analyzed theoretically by
the present authors in Ref.~\cite{ours}.

In the present work we analyze static and 
spectral properties of single- and few-qubit lasers, 
focusing on the regime of strong qubit-resonator coupling realized in 
most circuit QED experiments. Using a Master equation 
approach we analyze the consequences of qubit-field correlations. 
In the strong coupling regime they have significant quantitative 
effects on the laser line-width. 
In addition for the case of few-qubit lasing, we evaluate the corrections due to 
qubit-qubit correlations.  They are strongest at the transition to the lasing regime 
but yield only small corrections to the power spectrum. 
In the frame of the so-called  
``semi-quantum'' approximation \cite{mandel} we describe the 
qualitative differences between multi-atom lasers and 
superconducting micromasers; specifically we analyze the scaling of 
the lasing transition  and diffusion constant with the number of 
atoms. 

 The paper is organized as follows.
 We start discussing in the following Section the two experimental 
 realizations of superconduc\-ting micromasers reported in Refs.~\cite{astafiev} and \cite{ilichev03,grajcar}, respectively. In particular we describe how the population inversion in the qubit is created in the two examples.
In Section \ref{Model}, we formulate the theoretical model and  
derive the dynamical equations for the micromaser using a master equation approach. In Section \ref{Static} we review the theory of lasing, paying attention to effects which are usually ignored for conventional lasers 
but are prominent in single- or few-atom lasers in the strong coupling regime. 
We introduce the different approximation schemes.
Static properties of single-qubit lasers 
are presented, such as the average photon 
number and  the qubit-field and qubit-qubit correlations. 
We show explicitly that due to spontaneous emission in few-qubit lasers 
the sharp lasing threshold is replaced by a smooth, but still well localized transition to the lasing regime.

Next, in Section~\ref{Spectrum}, we analyze the spectral 
properties of superconducting micromasers. We discuss the effects of correlations between qubit and resonator on the phase diffusion process and we show that, beyond the lasing threshold, the linewidth grows with increasing coupling strength, thus deteriorating the lasing state. Finally, in 
section \ref{lfn} we analyze the scaling of the photon number and of the 
diffusion constant with the number of atoms. In addition we demonstrate how low-frequency noise leads to inhomogeneous broadening of the lasing peak.

\section{Inversion mechanisms in superconducting micromasers}
\label{Inversion}
\subsection{The SSET laser}\label{sect-SSET}
\begin{figure}[b]
\begin{center}
\includegraphics[width = 5 cm]{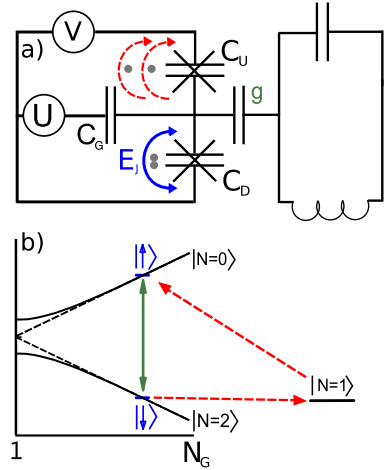}
\caption{a) The SSET consists of two superconducting leads
        coupled to an island via Josephson junctions with
        capacitance $C_{\rm U}$ and $C_{\rm D}$. 
        A transport voltage $V$ is applied, and the electrostatic energy
         of the island is tuned by the gate voltage $U$ via the capacitance $C_{\rm G}$. In addition, the SSET is capacitively coupled
         to an LC-oscillator with strength $g$.\\
      b) Energies of the charge states 
        $|N=0\rangle$ and $|N=2\rangle$ (dashed lines) 
        and the eigenstates $|\uparrow \rangle$ 
         and $|\downarrow\rangle$ (full lines).
         The energy of the odd charge state $|N=1\rangle$
      may be far from the other ones and 
          is drawn at an arbitrary position. 
         If the charge states $|N=0\rangle$ and $|N=2\rangle$ 
         are not close to degeneracy, Cooper pair tunneling
        is suppressed. For $N_{\rm G}>1$
         we have $\cos \frac{\xi}{2}>\sin \frac{\xi}{2}$,
        and the dominant quasiparticle transitions lead from $|\downarrow\rangle$
      to $|\uparrow\rangle$, as indicated by the dashed arrows.
      The capacitive coupling to the LC-oscillator
      creates an additional coupling between the states 
      $|\uparrow\rangle$ and $|\downarrow\rangle$,
      indicated by the vertical arrow.}
\label{fig:SystemandCycle}
\end{center}
\end{figure}

The ``SSET laser'' realized by Astafiev \emph{et al.} \cite{astafiev} 
consists of an SSET coupled capacitively to a microstripline resonator, as shown in Fig.~\ref{fig:SystemandCycle}a.  
The properties of the coupled system and the specific form of the Hamiltonian will be analyzed further in later Sections and in Appendix A. 
In the present Section we describe how a population inversion is created in a suitably biased SSET.

A superconducting single-electron transistor (SSET) consists of two superconducting leads
 coupled by tunnel junctions to a superconducting island.
 A gate voltage $U$ shifts the
 electrostatic energy of the island and controls,
 together with the bias voltage $V$, the current through the device.
 The Josephson coupling, $E_{\rm J}$,  allowing  for
 coherent Cooper pair tunneling through the junctions, is weak compared
 to  the superconducting energy gap $\Delta_{sc}$ and to the charging energy of the island, $E_C= e^2/2C$, 
 $C$ being the total island's
 capacitance. 
 In addition, quasiparticles can tunnel incoherently (with rate
 $\propto V/eR$, where  $R$ is the resistance of the tunnel junction)  when
 the energy difference between initial and final states
 is sufficient to create a quasiparticle excitation, i.e., when it
 exceeds twice the gap, $|\Delta E |\ge2\Delta_{sc}$.
 We denote by $N$ the number of excess charges on the island; it changes by $\pm 1$ in a single-electron tunneling process and by $\pm 2$ in a Cooper pair tunneling event. 
 At low temperatures for the conditions realized experimentally
 the number of accessible charge states of the island is  strongly reduced. For 
 the further calculations we can restrict our analysis to $N=0,1,2$.

 We assume the SSET to be tuned close to the Josephson quasiparticle (JQP) cycle,
 where the current is transported by a combination
 of Cooper pair tunneling through one junction and two consecutive quasiparticle tunneling events through the other junction.
 The parameters of the junctions are chosen asymmetrically.
 By changing the transport voltage $V$ and the gate voltage $U$, 
  we can tune to a situation, where resonant Cooper pair tunneling is strong
 across, say, the lower junction, while
 quasiparticle tunneling is strong across the upper one.
 Specifically, when the normalized ``gate charge" 
 $N_{\rm G}=C_{\rm G} U/e$ is approximately $1$,
    $N_{\rm G} \approx 1$,
 the charge states $|N=0\rangle$ and $|N=2\rangle$ are near degeneracy
 with respect to coherent Cooper pair tunneling across the lower junction.
Hence the eigenstates are
\begin{eqnarray}\label{eq:eigen}
  |\uparrow\rangle=\cos \frac{\xi}{2}\,|N=0\rangle+\sin \frac{\xi}{2}\,|N=2\rangle
  \nonumber\\
  |\downarrow\rangle=\sin \frac{\xi}{2}\,|N=0\rangle-\cos \frac{\xi}{2}\,
  |N=2\rangle
\end{eqnarray}
 where $\tan \xi=E_{\rm J}/\epsilon_{ch}$ with $\epsilon_{ch} = 4(N_{\rm G}-1) E_C$. Quasiparticle tunneling across the upper junction leads to transitions between the states $|\uparrow\rangle$ 
 and $|\downarrow\rangle$ and the odd charge state 
 $|N=1\rangle$. The transition rates are 
 \begin{eqnarray}\label{eq:RatesSSET}
  \Gamma_{\downarrow\rightarrow 1}=\Gamma_{1\rightarrow \uparrow}
  =\cos^2\left(\frac{\xi}{2}\right) I(V)\nonumber \\
  \Gamma_{\uparrow\rightarrow 1}=\Gamma_{1\rightarrow \downarrow}
  =\sin^2\left(\frac{\xi}{2}\right) I(V). 
 \end{eqnarray}
 The dependence on the relevant matrix elements and the energy gain $eV$
 can be lumped into the function $I(V)$, which is the normal current through the 
 junction at voltage $V$. Here we can assume that the relevant energy scale for
 each tunnel event is the applied voltage and neglect the smaller change of the energy 
 of the island.

 By choosing $\xi$ such that $\cos\frac{\xi}{2} > \sin\frac{\xi}{2}$, we can create a population inversion. In this case, the quasiparticle tunneling processes (\ref{eq:RatesSSET}) leading from $|\downarrow\rangle$ via $|1\rangle$ to $|\uparrow\rangle$ become stronger than the processes in  opposite direction (see fig. \ref{fig:SystemandCycle}b)).
 From the transition rates (\ref{eq:RatesSSET})
 we readily obtain the bare population inversion in the system,
 \begin{eqnarray}
 D_0= \frac{\rho_{\uparrow\uparrow}-\rho_{\downarrow\downarrow}}{\rho_{\uparrow\uparrow}+\rho_{\downarrow\downarrow}} = \cos\xi 
 \end{eqnarray}
where $\rho_{ii}$ is the population of the state $|i\rangle$ with $i=\uparrow,\,\downarrow$.

\subsection{The superconducting dressed-state laser}
\begin{figure}
\begin{center}
\includegraphics[width=7cm]
{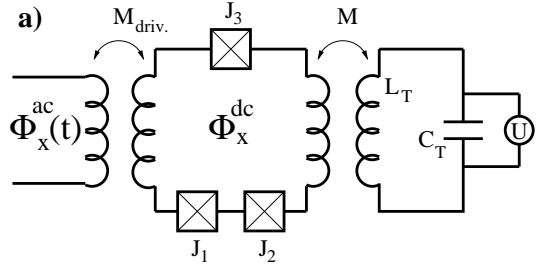}
\caption{In the setup of Ref.~\cite{ilichev03} an externally driven
three-junction flux qubit is coupled inductively to an LC
oscillator.} \label{fig:system}
\end{center}
\end{figure}

In this subsection we consider systems as investigated in Ref.~\cite{ilichev03,grajcar} and shown in Fig.~\ref{fig:system}.
Here, a flux qubit is strongly driven by $ac$-fields 
to perform Rabi oscillations. It is further coupled to a
low-frequency $LC$-oscillator.
In the strongly driven situation the physics is most conveniently described in the basis of ``dressed states" in the rotating frame \cite{PhysRev.188.1969}.  
The transformation to dressed states modifies the relaxation, excitation and decoherence rates as compared to the standard
results~\cite{Bloch_Derivation,Redfield_Derivation}. As a result, 
for blue detuning of the driving 
frequency compared to the resonant frequency
a population inversion is produced in the dressed state basis, which 
in turn can lead to lasing~\cite{Zakrzewski}.

To illustrate these effects we first consider the driven qubit (ignoring
the coupling to the resonator) coupled to a bath observable $\hat
X$,
\begin{eqnarray}
\label{eq:H_eigenbasis_NO} {H} & = & -\frac{1}{2}\, \Delta
E\,{\sigma}_{z}+\,\hbar\Omega_{\rm R0}\,
\cos\left(\omega_{d}t\right)\,{\sigma}_{x}\nonumber\\
  & & - \frac{1}{2} \left(b_x \sigma_x +
   b_y \sigma_y + b_z \sigma_z\right)\hat X +
H_{\rm bath}\ .
\end{eqnarray}
In the absence of driving, $\Omega_{\rm R0}=0$,
and for regular (i.e., smooth as function of the frequency) power spectra 
of the fluctuating bath observables we can proceed using Golden rule type 
arguments~\cite{Bloch_Derivation,Redfield_Derivation}.
The transverse noise, coupling to $\sigma_x$ and $\sigma_y$, is responsible for  relaxation and excitation processes with rates
\begin{eqnarray} \label{eq:Gamma ar}
\Gamma_{\downarrow} & = & \frac{|b_\perp|^2}{4\hbar^2}
\langle\hat{X} ^{2}\rangle_{\omega=\Delta E} \nonumber \\
\Gamma_{\uparrow} & = & \frac{|b_\perp|^2}{4\hbar^2}
\langle\hat{X} ^{2}\rangle_{\omega=-\Delta E} \, ,
\end{eqnarray}
while longitudinal noise, coupling to $\sigma_z$, produces a pure dephasing with rate
\begin{eqnarray} \label{eq:Gamma_*}
\Gamma_{\varphi}^* & = &  \frac{|b_z|^2}{2\hbar^2} S_{X}(\omega =0) \ .
\end{eqnarray}
Here $b_\perp \equiv b_x + i b_y$, and we introduced the ordered
correlation function
$\langle\hat{X} ^{2}\rangle_{\omega}\equiv\int dt\ e^{i\omega
t}\langle\hat{X} (t)\hat{X} (0)\rangle$,
as well as the power spectrum, i.e., the symmetrized correlation
function, $S_{X}(\omega)\equiv
(\langle\hat{X} ^{2}\rangle_{\omega}+\langle\hat{X} ^{2}\rangle_{-\omega})/2$.
The rates (\ref{eq:Gamma ar}) and (\ref{eq:Gamma_*}) also define the relaxation rate $1/T_1=
\Gamma_1= \Gamma_{\downarrow}
+\Gamma_{\uparrow} $ and the total dephasing rate $1/T_2 =\Gamma_
\varphi= \Gamma_1/2  + \Gamma_{\varphi}^*$ which appear in the Bloch
equations for the qubit.

To account for the driving with frequency $\omega_d$ it is convenient
to transform to the rotating frame via a unitary transformation $U_r =
\exp{(-i\omega_d\, \sigma_z t/2)}$. Within rotating-wave approximation (RWA) the transformed Hamiltonian
reduces to
\begin{eqnarray}
\label{eq:H_Rotating}
\tilde H &= &\frac{1}{2}\hbar \Omega_{\rm R0}\, {\sigma}_{x}+ \frac{1}{2}\hbar  \delta
\omega \, {\sigma}_{z} \nonumber \\
 &&-\frac{1}{2}\left[b_z{\sigma}_{z}+ b_\perp e^{i\omega_d t}\sigma_-
+ b_\perp^* e^{-i\omega_d t}\sigma_+ \right]\hat X  + H_{\rm bath} \ ,
\end{eqnarray}
with detuning $\delta\omega\equiv\omega_{d}-\Delta E/\hbar$.
The RWA cannot be used in the second line of
(\ref{eq:H_Rotating}) since the fluctuations
$\hat X $ contain potentially frequencies close to $\pm \omega_d$, which can compensate  fast oscillations.
Diagonalizing the first two terms of (\ref{eq:H_Rotating})
one obtains
\begin{eqnarray}
\label{eq:H_Rotating_diag}
\tilde H &=& \frac{1}{2}\hbar \Omega_{\rm R}\,\sigma_z+H_{\rm bath}
\nonumber \\
&&-\left[\frac{\sin\beta}{2}\, b_{z}+\frac{\cos\beta}{4} \,(b_\perp^*\,
e^{-i\omega_d t} + b_\perp\, e^{i\omega_d t}) \right]\,\sigma_z \, 
\hat X 
\\
&&-\Big\{\Big[\frac{\sin\beta+1}{4}\,b_\perp^*\, e^{-i\omega_d t} +
\frac{\sin\beta-1}{4}\,b_\perp \,e^{i\omega_d t} -\frac{\cos\beta}{2} \,b_z
\Big]\,\sigma_+\, \hat X + {\rm h.c.}\Big\}\ , \nonumber
\end{eqnarray}
where the full Rabi frequency is
$\Omega_{\rm R}=\sqrt{\Omega_{\rm R0}^{2}+\delta\omega^{2}} $, and the detuning determines the parameter $\beta$ via
\begin{eqnarray}
\tan\beta= \delta\omega/\Omega_{\rm R0} \, .
\end{eqnarray}

From here Golden-rule arguments lead to the relaxation and excitation
rates in the rotating frame as well as the "pure" dephasing rate~\cite{Saclay_Karlsruhe}
\begin{eqnarray} 
\tilde\Gamma_{\downarrow}  & \approx &
\frac{b_{z}^{2}}{4\hbar^{2}}\cos^{2}\beta \,
\langle\hat{X} ^{2}\rangle_{\Omega_{\rm R}} \nonumber\\
&& + \frac{|b_\perp|^2}{16\hbar^{2}} 
\left[\left(1-\sin\beta\right)^2
\langle\hat{X} ^{2}\rangle_{\omega_{d} +\Omega_{\rm R}}+\left(1+\sin\beta\right)^2
\langle\hat{X} ^{2}\rangle_{-\omega_{d}+\Omega_{\rm R}}\right]\
\nonumber\\
\tilde\Gamma_{\uparrow}  & \approx &
\frac{b_{z}^{2}}{4\hbar^{2}}\cos^{2}\beta \,
\langle\hat{X} ^{2}\rangle_{-\Omega_{\rm R}} 
\nonumber\\
&&+ \frac{|b_\perp|^2}{16\hbar^{2}} 
\left[\left(1-\sin\beta\right)^2
\langle\hat{X} ^{2}\rangle_{-\omega_{d}-\Omega_{\rm R}} +
\left(1+\sin\beta\right)^2
\langle\hat{X} ^{2}\rangle_{\omega_{d}-\Omega_{\rm R}}\right]\ ,
\nonumber%
\label{eq:Gamma_down}
\\
\label{eq:Gamma_puredeph}
\tilde\Gamma_{\varphi}^* & \approx &
\frac{b_{z}^{2}}{2\hbar^{2}}\sin^{2}\beta \, S_{X}(\omega=0)
+\frac{|b_\perp|^2}{4\hbar^{2}} \cos^{2}\beta  \, S_{X}(\omega_{d}).
\end{eqnarray}
We note the effect of the frequency mixing. In addition, due to the
diagonalization the effects of longitudinal and transverse noise on
relaxation and decoherence get mixed. We further note that
the rates also depend on the fluctuations' power spectrum at the
Rabi frequency, $\langle\hat{X} ^{2}\rangle_{\pm\Omega_{\rm R}}$.

For a sufficiently regular power spectrum of the fluctuations at frequencies
$\omega \approx \pm \Delta E/\hbar$ 
we can ignore the effect of detuning and the small
shifts by $\pm \Omega_{\rm R}$ as compared
to the high frequency $\omega_{d}\approx \Delta E/\hbar$.
We further assume that $\Omega_{\rm R}\ll k  T /\hbar$.
In this case we find the simple relations
\begin{eqnarray}
\label{mod}
\tilde \Gamma_\uparrow & = &
\frac{(1+\sin \beta)^2}{4}\, \Gamma_\downarrow
+ \frac{(1-\sin \beta)^2}{4}\, \Gamma_\uparrow
+ \frac{1}{2}\cos^2 \beta \, \Gamma_\nu \ , \nonumber \\
\tilde \Gamma_\downarrow & = &
\frac{(1-\sin \beta)^2}{4}\, \Gamma_\downarrow
+ \frac{(1+\sin \beta)^2}{4}\, \Gamma_\uparrow
+ \frac{1}{2}\cos^2 \beta \, \Gamma_\nu \ , \nonumber \\
\tilde\Gamma_{\varphi}^*   &=& \sin^2\beta\, \Gamma_{\varphi}^*+
\frac{\cos^{2}\beta}{2}(\Gamma_\downarrow+\Gamma_\uparrow)\ ,
\end{eqnarray}
where the rates in the lab frame are given by Eqs.~(\ref{eq:Gamma ar},\ref{eq:Gamma_*})
and the new rate
\begin{eqnarray}
\Gamma_\nu  \equiv  \frac{1}{2\hbar^2} \,b_{z}^{2} \,S_{X}(\Omega_{\rm R})
\end{eqnarray}
depends on the power spectrum at the Rabi frequency.

To proceed we concentrate on the most relevant regime.
At low temperatures, $k_{\rm B}T \ll \Delta E
\approx \hbar\omega_d$, we can neglect $\Gamma_\uparrow$ as it is
exponentially small. We also assume that 
$\Gamma_\nu$ can be neglected as compared to $\Gamma_\downarrow$,
which is justified, e.g., when the qubit is tuned close to the symmetry 
point where $b_z \ll |b_\perp |$. Since the rate $\Gamma_\nu$ depends on the noise power spectrum at the frequency $\Omega_{\rm R}$, which is usually higher that the 
frequency range of the $1/f$ noise, the latter does not change the situation.
Thus we neglect $\Gamma_\nu$ and we are left with
\begin{eqnarray} \label{eq:Gamma_down_res}
\tilde\Gamma_{\downarrow/\uparrow} \approx
\frac{\left(1\mp\sin\beta\right)^2}{4}
\,\Gamma_
\downarrow\quad  ,\quad \tilde\Gamma_{\varphi}^*  \approx
\frac{\cos^{2}\beta}{2}\,\Gamma_
\downarrow\ .
\end{eqnarray}

The ratio of up- and down-transitions depends on the detuning and can
be expressed by an effective temperature. Right on
resonance, where $\beta=0$, we have
$\tilde\Gamma_{\uparrow}=\tilde\Gamma_{\downarrow}$, corresponding to
infinite
temperature or a classical drive.  For
``blue" detuning, $\beta > 0$, we find $\tilde\Gamma_{\uparrow} >
\tilde\Gamma_{\downarrow}$, i.e., {\it negative temperature}.
This leads to a population inversion of the qubit, which is the basis
for the lasing behavior which will be described below.

In a more careful analysis, paying attention to the  small frequency shifts by $\pm \Omega_{\rm R}$, we obtain for $\beta =0$
\begin{equation}
\frac{\tilde\Gamma_\downarrow}{\tilde\Gamma_\uparrow} =
\frac{\langle\hat{X} ^{2}\rangle_{\omega_{d}+\Omega_{\rm R}}}
{\langle\hat{X} ^{2}\rangle_{\omega_{d}-\Omega_{\rm R}}} \, .
\end{equation}
For example, for Ohmic noise and low bath temperature this reduces 
to $\tilde\Gamma_\downarrow/\tilde\Gamma_\uparrow
\approx 1 + 2\Omega_{\rm R}/\omega_d$, which corresponds to an effective
temperature of order $2\hbar\omega_d/k_{\rm B}\approx 2\Delta
E/k_{\rm B}$, which by assumption is high but finite. The infinite
temperature threshold is crossed toward negative temperatures
at weak blue detuning when the condition
\begin{equation}
\frac{(1+\sin\beta)^2}{(1-\sin\beta)^2} \sim 1+\frac{2\Omega_{\rm R}}
{\omega_d}\
\end{equation}
is satisfied. We note that all qualitative features are well reproduced by the 
approximation (\ref{eq:Gamma_down_res}).

\begin{figure}
\begin{center}
\vspace*{0.5cm}
\includegraphics[width=0.65\columnwidth]{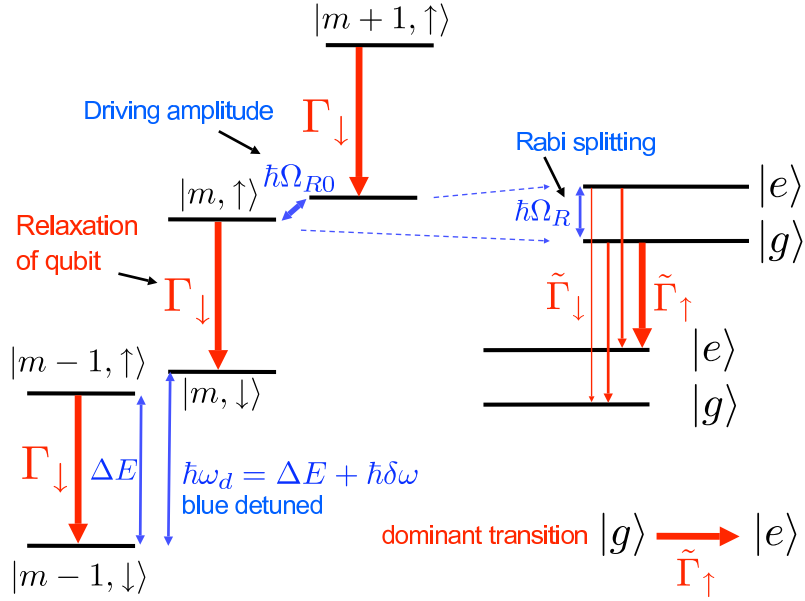} 
\caption{\rm Relaxation rates in the basis of the dressed states.
The left staircase denotes the eigenstates of the undriven qubit, $|\uparrow\rangle,|\downarrow\rangle$ and 
 the (quantized) driving field, $|m\rangle$, before the driving is switched on, i.e., for $\Omega_{\rm R0}=0$. 
The Hamiltonian in this basis is obtained from (\ref{eq:H_eigenbasis_NO}) by 
replacing $\hbar \Omega_{\rm R0}\cos\omega_d t$
with $\lambda (d^\dag +d)$, where $\lambda$ is the coupling  between the qubit and  driving field, and
$d, d^\dag$ are the annihilation and creation operators of the driving field.
The right staircase stands for the dressed states of a driven qubit near resonance, 
obtained by diagonalizing the corresponding $2\times2$ 
Hamiltonian. This yields 
$|g\rangle=\cos\left(\frac{\pi}{4}-\frac{\beta}{2}\right)|m,\uparrow\rangle
+\sin\left(\frac{\pi}{4}-\frac{\beta}{2}\right)|m+1,\downarrow\rangle$ and 
$|e\rangle=-\sin\left(\frac{\pi}{4}-\frac{\beta}{2}\right)|m,\uparrow\rangle
+\cos\left(\frac{\pi}{4}-\frac{\beta}{2}\right)|m+1,\downarrow\rangle$. 
The bare Rabi 
frequency is $\Omega_{\rm R0} \approx \lambda\sqrt{\bar m}$, where $\bar m$ is the 
average photon number in the coherent (classical) driving field. In the lab frame at low temperature only the relaxation rate $\Gamma_\downarrow$ needs to be considered.
However, in the dressed states basis the dominant rate $\tilde{\Gamma}_{\uparrow}$ leads from the state $|g\rangle$
to the state $|e\rangle$, thus creating a population inversion.}
\label{fig:mollow}
\end{center}
\end{figure}

To illustrate the calculations outlined above and the mechanism creating 
the population inversion for blue
detuning we show in Fig.~\ref{fig:mollow} the level structure, 
i.e., the formation of dressed states, of a near-resonantly driven qubit.
For the purpose of the present discussion we assume that also the driving field is quantized. This level
structure was described first by Mollow~\cite{PhysRev.188.1969}.
The picture also illustrates how for blue detuning
a pure relaxation process, $\Gamma_\downarrow$, in the laboratory frame 
predominantly leads to an excitation process, $\tilde\Gamma_\uparrow$, in the rotating frame, thus creating a population inversion in the basis of ``dressed states". 

If this effective inverted two-state system is coupled to an oscillator, a lasing state is induced. In Ref.~\cite{Zakrzewski} it was proposed to couple the oscillator to the dressed states 
belonging to the neighboring doublets (see Fig.~\ref{fig:mollow}). Then, to be in resonance with 
the pair of dressed states with population inversion the oscillator frequency should satisfy $\omega_0 = \omega_d + \Omega_{\rm R}$. 
As $\omega_d \sim \Delta E$ and $\Omega_{\rm R}\ll \Delta E$ 
this can work for a high-frequency resonator approximately in resonance with the qubit, $\omega_0 \approx \Delta E$. 
In contrast, in Ref.~\cite{hauss08} a different situation was considered
where the oscillator was coupled to the dressed states belonging to the same doublet.
The resonance condition then reads $\omega_0 = \Omega_{\rm R}$,  and the lasing can be reached for an oscillator much slower than the qubit, 
$\omega_0\ll \Delta E$, which is the situation realized in Ref.~\cite{ilichev03}. 
An additional complication arises at the symmetry point of the qubit, since there the 
single-photon coupling between the oscillator and the doublet of the dressed states vanishes.
Then, two-photon processes become relevant with the resonance condition 
$2\omega_0 = \Omega_R$~\cite{hauss08}.

\section{Modelling the single- or few-qubit laser} \label{Model}

We consider a single-mode quantum resonator coupled to $N_a$ qubits  (labelled by $\mu$). In the absence of dissipation, in the rotating wave approximation, the dynamics of the system is described by  the  Tavis-Cummings Hamiltonian \cite{tavis}:
\be H_{\rm TC}= \! \hbar \o_0 a^{\dagger} a +\frac{1}{2}\hbar 
\epsilon\sum_{\mu}\s^{\mu}_z + \hbar g\sum_{\mu} \lf( \s^{\mu}_+ a + 
\s^{\mu}_- a^{\dagger}\rg).\ee
Here we introduced, apart from the photon annihilation and creation operators, 
$a$ and $a^{\dag}$, the Pauli matrices acting on the single-qubit eigenstates 
$\s^{\mu}_z=\lf|\uparrow_{\mu}\rg>\lf<\uparrow_{\mu}\rg|-\lf|\downarrow_{\mu}\rg>\lf<\downarrow_{\mu}\rg|$, 
$\s^{\mu}_{+}=\lf|\uparrow_{\mu}\rg>\lf<\downarrow_{\mu}\rg|$, and $\s^{\mu}_{-}=\lf|\downarrow_{\mu}\rg>\lf<\uparrow_{\mu}\rg|$. Including both  
resonator and qubit dissipation the total Hamiltonian becomes
\be\label{H} H = H_{\rm TC}+(a + a^{\dagger}) X_a 
+\sum_{\mu}\lf(X^{\mu}_z \s^{\mu}_z + X^{\mu}_{+} \s^{\mu}_+ + 
X^{\mu}_{-} \s^{\mu}_- \rg) +H_{\rm bath}.
 \ee
Dissipation is modeled by assuming that the oscillator and  the qubits interact 
with noise operators, $X_a$ and $X^{\mu}_z$, $X^{\mu}_+$, 
$X^{\mu}_{-}$, belonging to independent baths with Hamiltonian 
$H_{\rm bath}$ in thermal equilibrium \cite{Gardiner}. The noise coupling 
longitudinally to the qubits,  $X^{\mu}_z \s^{\mu}_z$, is 
responsible for the qubits' pure dephasing. 

In this section and beyond we do not describe anymore the detailed mechanism creating  the population inversion in the qubits, which is necessary to obtain lasing.
Rather we introduce it by assuming that the effective temperature fixing the ratio of excitation and relaxation rates of the qubits is negative. (In the same spirit the transition rates $\Gamma$ appearing below are those of the effective two-level system, even if they refer to transition between dressed states, for which the rates were denoted above by $\tilde\Gamma$.)
Possible deviations from the Tavis-Cummings  oscillator-qubit coupling used 
in Eq. (\ref{H}) are discussed in Appendix A. 

The dynamics of a  single- or few-qubit 
laser can be analyzed in the frame of a master equation approach, as discussed by several authors \cite{hauss08,Ashhab,mu,briegel,rodrigues07-1}. In the Schr\"odinger picture the master equation for the reduced density 
matrix $\rho$ of the qubits and the oscillator reads 
 \begin{equation} \label{ME}
\dot \rho=-\frac{i}{\hbar}\left[H_{\rm TC}, \rho\right]
+L_{\rm Q}\, \rho+L_{\rm R} \, \rho \, .
\end{equation}
The Liouville operators $L_{\rm R}$ and $L_{\rm Q}$ describe the resonator's and qubits' dissipative processes. For Markovian processes it is sufficient to approximate them by Lindblad forms,
 \begin{eqnarray}\label{damp} 
L_{\rm R} \, \rho=& &\frac{\kappa}{2}\left[(N_{\rm th}+1)
\left( 2a \rho a^{\dagger} -a^{\dagger} a\rho -\rho 
a^{\dagger}a \right)\right.\nn
\\
& &+\left. N_{\rm th} \left( 2 a^{\dagger} 
\rho a-a a^{\dagger}\rho -\rho aa^{\dagger} \right)\right]
\end{eqnarray}
and 
 \begin{eqnarray}\label{L_Q^R}
L_{\rm Q} \, \rho =& &\sum_{\mu}\lf[\frac{\Gamma_{\varphi}^*}{2} 
\left(\sigma^{\mu}_z\rho\sigma^{\mu}_z - 
\rho\right)+\frac{\Gamma_{\downarrow}}{2}\left(2 \sigma^{\mu}_- \rho 
\sigma^{\mu}_+ -\rho\sigma^{\mu}_+\sigma^{\mu}_- 
-\sigma^{\mu}_+\sigma^{\mu}_-\rho 
\right)\rg.\nn\\
& &+\lf.\frac{\Gamma_{\uparrow}}{2}\left(2 \sigma^{\mu}_+ 
\rho\sigma^{\mu}_- -\rho\sigma^{\mu}_-\sigma^{\mu}_+ - 
\sigma^{\mu}_-\sigma^{\mu}_+\rho \right)\rg] .
\end{eqnarray}
The dissipative evolution of the system depends on the excitation, relaxation, and pure dephasing rates of the qubits, $\Gamma_{\uparrow}$, $\Gamma_{\downarrow}$ and $\Gamma_{\varphi}^*$, as well as on the bare damping rate of the resonator, $\kappa$, and on the thermal photon number $N_{\rm th}$.
 
For later purposes, we also introduce the rate $\G_1=\Gamma_{\downarrow}+\Gamma_{\uparrow}$, which is the 
sum of excitation and relaxation rates, and $\G_\vf=\Gamma_1/2+\G_{\vf}^*$, the total dephasing rate, incl. the ``pure dephasing" due to longitudinal noise described by $\G_{\vf}^*$. In contrast to relaxation and excitation 
processes, pure dephasing arises due to processes with no energy 
exchange between qubit and environment and thus does not affect the 
populations of the two qubit states. The parameter $D_0 = 
(\Gamma_{\uparrow}-\Gamma_{\downarrow})/\Gamma_1$ denotes the 
stationary qubit polarization in the absence of the resonator. In 
the present case, since we assume a negative temperature of the 
qubit baths and a population inversion, we have $D_0>0$.

The master equation (\ref{ME}) allows us to determine completely the quantum state of the system. However, its full solution is 
numerically demanding  in the experimental regime of parameters due to the high number of photons in the resonator  (of the order of $10^2$ or higher for a single-qubit laser). For this reason, we will use, whenever possible, different approximation schemes to calculate the physically 
relevant quantities.

\section{Approximations and  static properties} \label{Static}

To describe the single- or few-qubit laser in the strong coupling regime 
we start from the master equation (\ref{ME}) for the density matrix. 
In some cases we find that approximate analytical results, 
which are presented in this section, are sufficient.
In general, however, we rely on a numerical solution. 

From Eq. (\ref{ME}) we obtain the following equations for the  average 
photon number $\la n\ra$, the qubit polarization $\la\s^{\mu}_z\ra$ 
and the product $\la\sigma^{\mu}_+ a\ra$,
\begin{eqnarray}
   & & \frac{d}{dt} \langle \s_z^\mu\ra=- 2 i g \lf( \la\s^{\mu}_{+} a\ra - \la\s^{\mu}_{-} a^{\dagger}\ra \rg)  - \G_1 (\la\sigma^{\mu}_z\ra - 
   D_0) \phantom{\sum_{\mu}},\nonumber\\
   & & \frac{d}{dt}\langle  n \rangle = i g \sum_{\mu} \left( \langle \sigma^{\mu}_+ a \rangle - 
   \langle \sigma^{\mu}_- a^{\dagger} \rangle \right) - \kappa \left( \langle n \rangle - N_{\rm th} \right), \label{eqn}\nonumber\\
   & & \frac{d}{dt} \langle \sigma^{\mu}_+ a \rangle = \left(  i \Delta-\gamma  \right) 
   \langle \sigma^{\mu}_+ a \rangle - i g \langle \sigma^{\mu}_z n \rangle - i g \sum_{\nu} \langle \sigma^{\mu}_+ \sigma^{\nu}_- \rangle \, . 
   \label{eqs+a}
\end{eqnarray}
Here we introduced the detuning $\D=\epsilon-\o_0$ and the total 
dephasing rate $\g=\Gamma_\vf+\frac{\k}{2}$.

 In the stationary limit, after isolating the correlations between 
different qubits by writing $\langle \sigma^{\mu}_+ 
\sigma^{\nu}_- \rangle = \delta_{\mu\nu}(1+\langle \sigma_z^{\mu} 
\rangle)+(1-\delta_{\mu\nu})\langle \sigma^{\mu}_+ \sigma^{\nu}_- 
\rangle$, we derive from the previous equations  the following two exact 
relations   between four quantities: the average qubit polarization  $\la 
S_z(t)\ra$ with $S_z \equiv \frac{1}{N_a}\sum_{\mu}\s_z^{\mu}$, the 
photon  number  $\la n(t) \ra$,  and  the correlators $\la n S_z \ra$ and $C_{QQ}= \frac{1}{N_a} \sum_{\mu\neq\nu} \langle 
\sigma^{\mu}_+ \sigma^{\nu}_- \rangle$, 
 \bea
     \la n \ra & =& N_{\rm th} + \frac{2 g^2 N_a}{\k} \frac{\g}{\g^2 + \D^2}
\lf[ \la S_z n \ra + \frac{1}{2}(\la S_z \ra + 1)+C_{QQ}\rg],  \, \nn\\
     \la S_z \ra & = & D_0 - \frac{4 g^2}{\G_1} \frac{\g}{\g^2 + \D^2}
 \lf[ \la S_z n \ra + \frac{1}{2}(\la S_z \ra + 1)+C_{QQ}\rg]\,.\label{stat}
\eea 
If two of them are known, e.g., from a numerical solution of the 
master equation, the other two can be determined. 

\subsection{Semi-quantum model}
 Factorizing the 
correlator, $\la S_z n \ra \approx \la S_z \ra\la  n \ra$, on the 
right-hand side of Eqs.~(\ref{stat}) and  neglecting the qubit-qubit 
correlations, $C_{QQ}\simeq0$, we reproduce results known in 
quantum optics as ``semi-quantum model" \cite{mandel}.
This 
approximation  yields a quadratic equation for the scaled average 
photon number (per qubit) $\tilde{n} = \langle n \rangle/N_a$,
\begin{eqnarray}
   & & \tilde{n}^2 + \left( \tilde{n}_0 - \frac{\Gamma_1 D_0}{2 \kappa} - \frac{N_{\rm th}}{N_a} + \frac{1}{2 N_a} \right) \tilde{n} \nonumber \\
   & & \qquad \quad - \left( \frac{N_{\rm th} \tilde{n}_0}{N_a} + \frac{N_{\rm th}}{2 N_a^2} + \frac{\Gamma_1}{4 \kappa}\frac{D_0+1}{N_a} \right) = 0,
\end{eqnarray}
which depends on the parameter $\tilde{n}_0 = \frac{\Gamma_1 
\gamma}{4 g^2 N_a} \left( 1+\frac{\Delta^2}{\gamma^2} \right)$. This 
equation has always one positive solution $\tilde{n}>0$.

\subsection{Semiclassical approach}

Before continuing the analysis of the properties of the 
semi-quantum solution, for sake of comparison, we recall the 
standard semiclassical results.  In this
approximation the operator $a$ is treated  as a classical 
stochastic variable, $\alpha$. After adiabatic 
elimination of the qubits' degrees of freedom, i.e., assuming 
$\Gamma_{\varphi} \gg \kappa/2$, one obtains a classical Langevin 
equation for $\alpha$,  
\begin{equation}
\dot{\alpha} = - \left[ \frac{\kappa}{2} - \frac{g^2 N_a 
}{\Gamma_{\varphi}-i\Delta}s_z^{\rm st} \right] \alpha + 
\xi(t)\,. \label{SCLangevin}
\end{equation}
Here $\xi(t)$ is a classical Langevin force due to thermal noise, 
 $\langle \xi(t) \xi^*(t') \rangle = \kappa N_{\rm th} 
\delta(t-t')$, and $s_z^{\rm st} = D_0/(1+\left|\alpha\right|^2/(\tilde{n}_0N_a))$ denotes the stationary  qubits' polarization. 

In order to obtain an expression for the average photon number 
$\langle n \rangle = \langle \left| \alpha \right|^2 \rangle$ 
we rewrite the Langevin equation as 
$\dot{\alpha} = -f(\left|\alpha \right|^2) \alpha + \xi(t)$ 
and approximate $\langle f(\left| \alpha \right|^2) \cdot \left| \alpha \right|^2 \rangle \approx f(\langle n \rangle) \cdot \langle n \rangle$.
Thus we arrive at the equation 
$\frac{d}{dt} \langle n \rangle = -2\mathrm{\rm Re} \lbrace f(\langle n \rangle) \rbrace \cdot \langle n \rangle + \kappa N _{\rm th}$, from which we obtain 
in the steady state a quadratic equation for the scaled photon number $\tilde{n} = \la n \ra / N_a$,
\begin{equation}
 \tilde{n}^2 + \left( \tilde{n}_0 - \frac{\Gamma_1 D_0}{2 \kappa}  - 
 \frac{N_{\rm th}}{N_a} \right) \tilde{n} - \frac{N_{\rm th} \tilde{n}_0}{N_a} = 0 
\end{equation}
In the low-temperature limit, $N_{\rm th}\sim 0$, the semiclassical 
results can be rewritten in the simple form, $\tilde{n}^2 + \left( \tilde{n}_0 - \Gamma_1 D_0/2 \kappa \right) \tilde{n} \approx 0$, with $\tilde{n}_0 = \frac{\Gamma_1 
\Gamma_{\varphi}}{4 g^2 N_a} \left( 1+\frac{\Delta^2}{\Gamma_{\varphi}^2} \right)$. This gives the well-known threshold condition $D_0 > \kappa\Gamma_{\varphi} 
/ (2g^2 N_a)$ for the lasing state.

\subsection{Comparison of the different approaches}

Different from the semiclassical picture,  the semi-quantum model 
includes  the effects of spontaneous emission processes, 
described  by the term proportional 
to $(\la S_z\ra+1)$ in Eqs.~(\ref{stat}). 
Spontaneous emission is 
responsible for the linewidth of the lasers, and, as noticed in 
Ref.~\cite{mu}, due to the low photon number, spontaneous emission 
is especially relevant for the dynamics of 
single-atom lasers. 

\begin{figure}
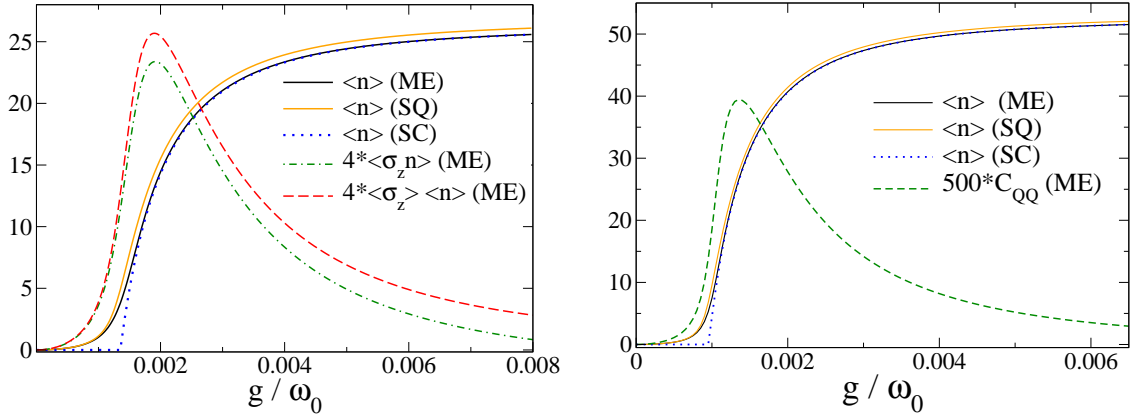

\begin{center}
\vspace*{0.5cm}
\includegraphics[width=7.4cm]{ncompareNa1.eps}
\hspace{0.3cm}
\includegraphics[width=7.0cm]{QCor2.eps}
\caption{Left panel: average photon number $\la n\ra$ in the resonator and qubit-field correlations for a single-qubit-laser. The photon number is calculated using the Master equation (ME, solid black line), the semi-quantum (SQ, solid orange (light grey) line), and the semiclassical approximation (SC, dotted line). The dot-dashed and dashed lines show the average values $\langle \sigma_z n \rangle$ and $\langle \sigma_z \rangle \langle n \rangle$. Right panel: average photon number $\la n\ra$  in the resonator and qubit-qubit correlations $C_{QQ}$ (dashed line) for a two-qubit-laser. The photon number was calculated using the Master equation (solid black line), the semi-quantum (solid orange (light grey) line), and the 
semiclassical approximation (dotted line). The  parameters are  $\epsilon = \o_0$, $\G_1/\o_0 = 0.016$, $\G_\vf^*/\o_0 = 0.004$, $D_0 = 0.975$, $\k/\o_0 = 3 \cdot 10^{-4}$, and $N_{\rm th} = 0$.}
\label{figcomp}
\end{center}
\end{figure}

To illustrate the effect of spontaneous emission on 
the lasing transition and at the same time the quality of the semi-quantum 
approximation, we plot the photon number as a function of the coupling strength 
$g$ for $N_a=1$ (Fig. \ref{figcomp}, left panel) and $N_a=2$ (Fig. \ref{figcomp}, rigth panel).
The plots show the semi-quantum, semiclassical and Master equation results. We note that the semi-quantum approximation gives results in very good agreement with the Master equation. Moreover, both the semi-quantum and Master equation solution show  a smooth crossover  between the normal and the lasing regimes,  an effect which is due to spontaneous emission. 
While we cannot define a sharp threshold condition, we can still identify, 
even for a single-atom laser, a well localized transition region centered at the  threshold coupling predicted by the semiclassical approximation.

In the left panel of Fig.~\ref{figcomp} we also plot the qubit-oscillator correlator, $\la S_z n \ra$, and the factorized approximation, $\la S_z\ra \la n \ra$. For strong coupling, they differ significantly. However, as the good agreement between the semi-quantum approximation and the numerical solution of the Master equation demonstrates, the qubit-field correlations have only a weak effect on the average photon number. On the other hand, as we will see in the following section, the qubit-field correlations have an important effect on the spectral properties of the single-qubit laser. 
                
The right panel of Fig.~\ref{figcomp}, shows the qubit-qubit correlations $C_{QQ}$ for a two-qubit-laser. Similar as the qubit-field correlations, they are neglected in the semi-quantum approximation. Both correlations are maximum at the lasing transition,
but decay away from this point. The reason is that qubit-field and qubit-qubit correlations scale as $g^2/\Gamma_{\vf}^2$, thus they are small for weak coupling. On the other hand, they are proportional to the qubit inversion $\la S_z \ra$ and hence vanish rapidly above the transition.

\section{Spectral properties} \label{Spectrum}

In this section, we will study the spectral properties of single-qubit lasers. The emission spectrum $\hat{O}(\omega)$ is given by the Fourier transform of the correlation function $O(\tau) = \lim_{t\rightarrow\infty} \langle a^{\dagger}(t+\tau) a(t)\rangle$. As we will see, for typical circuit QED parameters, i.e., for strong coupling $g$, the semi-quantum approximation, in spite of giving a sufficient estimate of the stationary photon number, cannot be used for a quantitative study of spectral functions. We evaluate the correlation function by performing a time-dependent simulation of the master equation (\ref{ME}) using the method described in Ref. \cite{briegel}. This method is numerically demanding, especially when we consider lasing with more than one qubit, $N_a>1$. We will show that in resonance, $\epsilon=\omega_0$, the semi-quantum theory catches the most qualitative features, both below and above the transition to the lasing regime. We will use this method later in Section \ref{Scaling} to investigate the scaling of the spectral properties with the number of qubits $N_a$. 

\subsection{Spectral properties in the semi-quantum theory}

Similarly to Eqs. (\ref{eqs+a}) for the average values, we can derive equations for the laser and cross correlation functions, $O(\tau)$ and $G(\tau) = \lim_{t\rightarrow\infty} \langle \sigma_+(t+\tau) a(t)\rangle$. Assuming that the oscillator damping is much weaker than the qubits' dephasing, $\kappa/2 \ll \Gamma_{\varphi}$, which is usually satisfied in single-qubit lasing experiments, we obtain a single equation for the oscillator correlation function: $\frac{d}{d\tau} \langle a^{\dagger}(t+\tau) a(t) \rangle = \left(i \omega_0 -\kappa/2 \right) \langle a^{\dagger}(t+\tau) a(t) \rangle + g^2/\Gamma_{\varphi} \langle \sigma_z \rangle \langle a^{\dagger}(t+\tau) a(t) \rangle$. Thus the semi-quantum theory predicts an exponential decay of the correlation function $O(\tau)$, which corresponds to a Lorentzian shape of the emission spectrum,
\begin{equation}
   \hat{O}(\omega) = \frac{2 \kappa_d \langle n \rangle}{(\omega-\omega_0)^2+\kappa_d^2}, 
\end{equation}
where the width of the spectrum is given by the expression
\begin{eqnarray}
 \kappa_d & = & \frac{\kappa}{2}\frac{N_{\rm th}}{\langle n \rangle} + 
 \frac{g^2 N_a}{\Gamma_{\varphi}}\frac{(\langle S_z \rangle +1)}{2 \langle n\rangle}. \label{kdSQ} 
\end{eqnarray}
\subsection{Numerical investigation of the spectral properties}
\begin{figure}
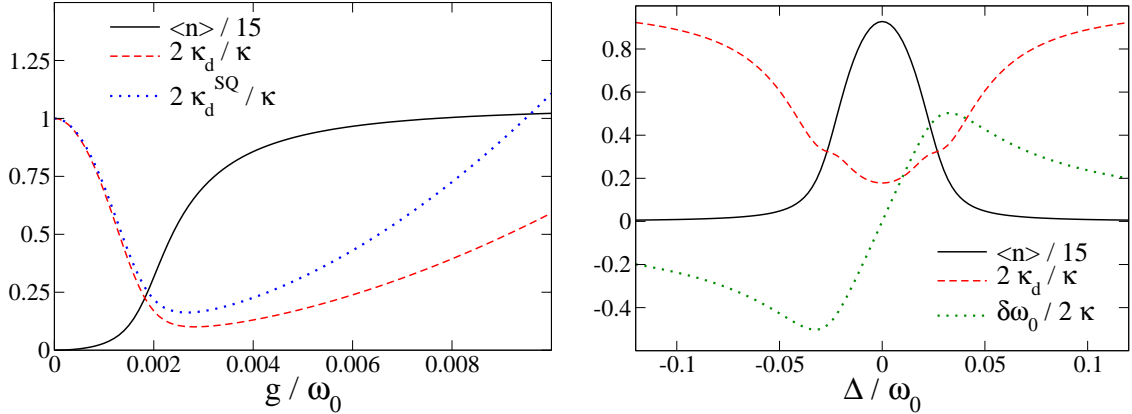

\begin{center}
\vspace*{0.6cm}
\includegraphics[width=7.3cm]{kdn.eps}
\hspace{0.2cm}
\includegraphics[width=7.2cm]{kdOmdn.eps}
\caption{Left panel: Phase diffusion constant and average photon number (solid line) as a function of the coupling strength $g$ for a single-qubit laser. The phase diffusion constant is calculated using the numerical simulation (dashed line) and the semi-quantum approximation (dotted line). Right panel: Phase diffusion constant (dashed line), frequency shift (dotted line) and photon number (solid line) as a function of the detuning for a fixed coupling strength $g/\omega_0=0.005$. In both plots, we used $\kappa/\omega_0 = 5\cdot10^{-4}$ for the bare damping rate of the resonator, other parameters as in Figure \ref{figcomp}.} \label{sq-hy}
\end{center}
\end{figure}

For the following discussion we focus on the case of a single qubit, $N_a=1$. In the left panel of Fig.~\ref{sq-hy} we plot the diffusion constant $\k_d$, as function of the coupling strength $g$, 
covering the whole range from below to above the transition, and compare it to the diffusion constant $\k_d^{\rm fac}$, obtained from the semi-quantum theory, i.e., by neglecting the qubit-field-correlations $\langle \sigma_z n \rangle - \langle \sigma_z \rangle \langle n \rangle$. 

Upon approaching the broadened lasing threshold from the weak coupling side we observe the linewidth narrowing characteristic for the lasing transition. However, above the transition, the linewidth increases again with growing coupling strength, thus deteriorating the lasing state. By comparing the semi-quantum approximation with the full solution of the master equation, we observe that qubit-oscillator correlations have a significant quantitative effect on the phase diffusion, leading to a \emph{reduction} of the linewidth by roughly a factor $1/2$, but they do not change the qualitative conclusions.   

Also in Fig.~\ref{sq-hy}, we note that in the transition region there 
is an ``optimal''  value of the qubit-oscillator coupling where the height of the spectral line, which is given by the ratio $\langle n \rangle / \kappa_d$ of the photon number and the linewidth, is maximum. This interesting feature is due to the fact that in single- and few-qubit lasers far above the lasing transition a increase of the coupling has little effect on the saturated photon number, but leads to an increase of the incoherent photon emission rate and the linewidth. 

When the qubit and the resonator are not in resonance, $\Delta \neq 0$, the emission spectrum is shifted with respect to the natural frequency $\omega_0$ of the resonator. This is shown in the right panel of figure \ref{sq-hy} where we plot the average photon number, the linewidth and the frequency shift $\delta \omega_0$ as functions of the detuning $\Delta$ in the strong coupling regime.

The numerical results for the linewidth $\kappa_d$ shown in this plot differ qualitatively from the results presented in our previous paper \cite{ours}, where we use a factorization scheme to obtain analytical expressions for the linewidth. Specifically, moving away from the resonance the linewidth is  \emph{increasing}, while the factorization predicted a decrease.  
On the other hand, the approximation based on the factorization yields results very similar to
 the numerical ones right on resonance, as well as in the far off-resonant situation and the weak-coupling regime.

\section{Discussion}\label{lfn} 
\subsection{Scaling in the semi-quantum approximation} \label{Scaling}
\begin{figure}
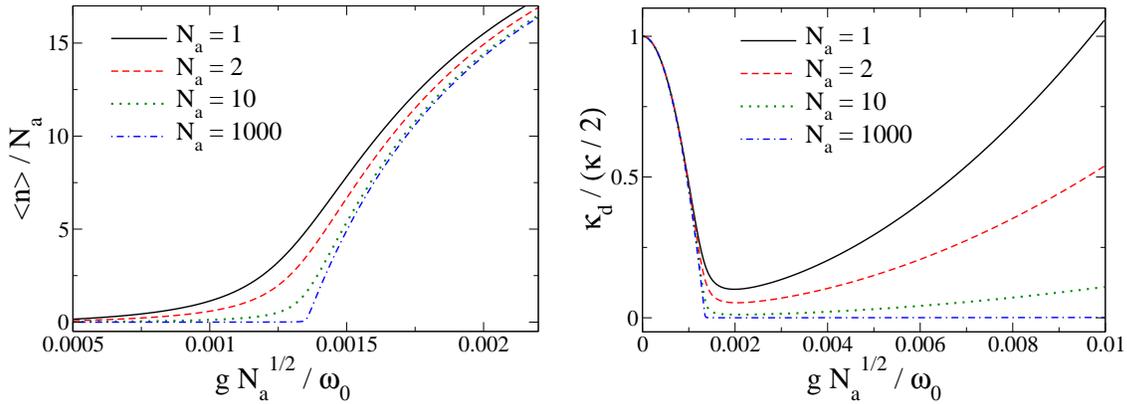

\begin{center}
\vspace*{0.5cm}
\includegraphics[width=7.1cm]{nscaling.eps}
\hspace*{0.2cm}
\includegraphics[width=7.3cm]{kdscaling.eps}
\caption{Average scaled photon number and phase diffusion constant 
as a function of the scaled coupling $g\sqrt{N_a}$. Other parameters as in Figure \ref{figcomp}.} \label{scaling}
\end{center}
\end{figure}
As discussed  in a quantum optics context in Refs.~\cite{mu,briegel}, various properties of single qubit masers are due  
to the fact that in these systems only one artificial atom (a 
\emph{microscopic} system from a thermodynamical point of view)  
interacts with the electromagnetic radiation.   To clarify the main 
differences between single qubit masers and conventional (many atom) 
lasers, we use the semi-quantum approximation to 
study the scaling  of the average photon number and of the phase 
diffusion with the number of atoms.

In Fig.~\ref{scaling}, we plot the scaled photon number $\langle n 
\rangle / N_a$ in the transition region and $\k_d$ versus the  
scaled coupling, $g\sqrt{N_a}$,  for different values of the number of qubits $N_a$. Plotted in these scaled forms, all curves have the same asymptotic behavior,  and the transition occurs at the same position. As expected, in  Fig.~\ref{scaling} (left panel) we observe that for low values of $N_a$, there is a smoothening of the lasing transition which is due to spontaneous emission processes and disappears in the large $N_a$ 
limit. In Fig.~\ref{scaling} (right panel) we show the scaling of 
the phase diffusion constant. Here the qubits' relaxation processes 
are responsible for the increase of the phase diffusion rate in the 
case of strong qubit-oscillator coupling for small $N_a$.

\subsection{Effect of the low-frequency noise}
The linewidth of order of 0.3MHz observed in  
Ref.~\cite{astafiev} is about one order of magnitude larger than 
what follows from our results (of the order of the 
Schawlow-Townes linewidth). Moreover, in the experiment, 
the emission spectrum shows a Gaussian rather than a Lorentzian shape. 
Both discrepancies can be explained if we note that the qubits' dephasing 
is mostly due to low-frequency charge 
noise, which cannot be treated within the Markov approximation used in the present analysis. 

However, low-frequency (quasi-static) noise can be taken into account by averaging the Lorentzian spectral line over different values of the energy splitting $\epsilon$ of the qubit \cite{falci}, or equivalently, over different values of the detuning $\Delta$ between qubit and oscillator. Assuming that these fluctuations are Gaussian distributed, with mean $\bar \D$  and width $\sigma$, such that $\Gamma_1>\sigma\gg \k_d$, we can neglect in the saturated limit the dependence of $\k_d$ and $\la n \ra$ on $\Delta$ and assume that the frequency shift $\d\o_0$ depends linearly on the detuning $\Delta$.
 
In the strong coupling regime, the shift of the emission spectrum is given, in a good approximation, by $\d\o_0\simeq \D\k/(2\Gamma_{\vf})$, which leads to a Gaussian line of width 
$\tilde\s \simeq \s\k/(2\Gamma_{\vf})$, where we remark that 
$\Gamma_\vf$ is the total \emph{Markovian} dephasing rate. In this way,
the linewidth observed in the experiment can be reproduced by a reasonable choice 
of $\s$ of order of 300 MHz. In the case in which $\sigma$ is larger 
than $\Gamma_1$,  the previous formula overestimates the linewidth 
since it does not take into account the decay of $\la n\ra$ below the lasing transition. In this case we can still perform the averaging numerically. 
In either case we note that in the presence of low-frequency noise, the linewidth is 
governed not by $\k_d$, but by $\d\o_0$.

\section{Conclusions}
We analyzed in detail the static and spectral 
properties of single- and few-qubit lasers. Our main conclusions are:\\
- As compared to a conventional laser setup with many atoms, which has a 
sharp transition to the lasing state at a threshold value of the coupling 
strength (or inversion), we find for a single- or few-qubit laser a smeared,
 but still well defined transition. Similarly, the decrease of the phase 
 diffusion strength when approaching the transition, i.e.,  
 the characteristic linewidth narrowing, is less sharp but still pronounced. \\
- Above the lasing trasition we observe for a single- or few-qubit laser a 
pronounced increase of the phase diffusion strength, 
which leads to a deterioration of the lasing state and a reduction of the hight of the laser 
spectrum.\\
- Low-frequency noise  strongly affects the linewidth  of the lasing 
peak,  leading to an inhomogeneous broadening.  In comparison, the 
natural laser linewidth due to spontaneous emission is negligible.
\section*{Acknowledgments}

We acknowledge fruitful discussions with O. Astafiev, J. Cole, A. Fedorov and F. Hekking. The work is part of the EU IST Project EuroSQIP.\\
\appendix

\section{Comment on two-photon processes} \label{Appendix1}

Here we briefly discuss the validity of the Jaynes-Cummings model introduced  in 
Section \ref{Model}, when applied to describe the SSET laser of 
Astafiev \emph{et al.}~\cite{astafiev}. As discussed in Section \ref{sect-SSET}, the SSET laser, schematically depicted in Fig. \ref{fig:SystemandCycle}, 
consists of  a  biased superconducting island  
coupled capacitively to a single-mode electrical resonator.  Under appropriate conditions 
only two charge states, corresponding to $N=0,2$  are relevant to the  dynamics of the device. In 
this basis the hamiltonian of the oscillator and the qubit can be written as
\be
   H = \hf \lf(\epsilon_{ch} \, \tau_z + E_J \tau_x\rg)+ \hbar \o_0 a^{\dagger} a - \hbar g_0 \tau_z \lf( a + a^{\dagger} \rg). \label{ChHamil}
\ee 
where the operators $\tau_x$ and $\tau_z$ are defined as: $\tau_z=\lf(\lf|N=2\rg>\lf<N=2\rg|-\lf|N=0\rg>\lf<N=0\rg|\rg)$ and  
$\tau_{x}=\lf(\lf|N=2\rg>\lf<N=0\rg|+\lf|N=0\rg>\lf<N=2\rg|\rg)$.
Rotating to the qubit's eigenbasis $\lf\{|\uparrow\ra,\, |\downarrow\ra\rg\}$, defined by Eqs. (\ref{eq:eigen}), 
we can recast the hamiltonian as follows:
\be
   H = \hf \epsilon \, \s_z + \hbar \o_0 a^{\dagger} a - \hbar g_0 \lf( \cos\xi\s_z - \sin\xi\s_x \rg)  \lf( a + a^{\dagger} \rg). \label{DiagHamil}
\ee 
The angle $\xi$  is defined as in Section \ref{sect-SSET}, $\tan\xi = \frac{E_{\rm J}}{\ve_{\rm ch}}$, and the qubit energy splitting, $\epsilon$, depends on the 
charging and Josephson energies, $\ve_{\rm ch}$ and $E_{\rm J}$:   $\epsilon = \sqrt{\ve_{\rm ch}^2+E_{\rm J}^2}$. %
In order to identify the one- and two-photon coupling strength, we  now 
apply a Schrieffer-Wolff transformation $U = e^{iS}$ with $S=i 
\frac{g_0 \cos \xi}{\o_0} \s_z \lf( a - a^{\dagger} \rg)$ and 
perform a perturbation expansion in the parameter $g_0/\o_0$. The 
transformed Hamiltonian, $ \tilde H= U^{\dag}HU$, thus becomes
\bea \tilde H  \simeq  \hf \epsilon \, \sigma_z + \hbar \omega_0 
a^{\dagger} a + \hbar g_1 \s_x \lf( a + a^{\dagger} \rg)  + \hbar g_2 i \s_y  \lf( a^2 - \lf( a^{\dagger} \rg)^2 
\rg) . \eea
Here we neglected terms of order $\lf(g_0/\o_0\rg)^3$ and  
introduced the two coupling constants $g_1 = - g_0 \sin \xi$ and $g_2 = \frac{2 g_0^2}{\o_0} \sin \xi \cos \xi $ for one-photon and two-photon transitions, respectively. For the parameters used in the experiment the coupling $g_2$ is roughly two orders of magnitude smaller than the one-photon coupling and  below the semiclassical threshold for the two-photon lasing, $ 
g_2^{\rm thr}=\sqrt{\k^2\G_\vf/(\G_1 D_0^2)}$ \cite{wang}. In the 
parameter regime explored in the experiments, the Hamiltonian  used 
in Eq. (\ref{H}) gives thus a good description of the dynamics of 
the system.


\section*{References}
\begin{thebibliography}{99}
\bibitem{blais04} A. Blais \emph{et al.},  Phys. Rev. A {\bf 69},   062320
(2004); R. Schoelkopf and S. Girvin, Nature {\bf 451},  664 (2008). 
\bibitem{ilichev03} E. Il'ichev \emph{et al.},  Phys. Rev. Lett. 
{\bf 91}, 097906 (2003).
\bibitem{wallraff04} A. Wallraff \emph{et al.}, Nature {\bf 431},  162
(2004).
\bibitem{chiorescu} I. Chiorescu \emph{et al.}, Nature  {\bf 431},  159 (2004).
\bibitem{johansson06}G. Johansson, L. Tornberg, and C. M. Wilson, Phys. Rev. B {\bf 74}, 100504 (2006).
\bibitem{sillanpaa}M. A. Sillanp\"a\"a, J. I. Park, and R. W. Simmonds, 
Nature {\bf 449}, 438 (2007); J. Majer \emph{et al.}, Nature {\bf 449},  443 (2007).
\bibitem{leek07} P.J. Leek \emph{et al.}, Science {\bf 318}, 1889 
(2007).
\bibitem{filipp}S. Filipp \emph{et al.}, Phys. Rev. Lett. {\bf 102}, 200402 (2009)); P.J. Leek \emph{et al.}, Phys. Rev. B {\bf 79}, 180511(R) (2009).
\bibitem{dicarlo}L. Di Carlo \emph{et al.}, Nature {\bf 460}, 240 (2009). 
\bibitem{hofheinz} M. Hofheinz \emph{et al.}, Nature {\bf 454}, 310 (2008); J. M. Fink  \emph{et al.}, Nature {\bf 454},  315 (2008). 
\bibitem{sillanpaa09}M. A. Sillanp\"a\"a \emph{et al.}, arxiv: 0904.2553 (2009).
\bibitem{astafiev} O. Astafiev \emph{et al.}, Nature {\bf 449},  588 (2007).
\bibitem{grajcar} M. Grajcar \emph{et al.}, Nature Physics {\bf 4},  612 (2008). 

\bibitem{haken} H. Haken, \emph{Laser Theory}, Springer, Berlin, 1984. 
\bibitem{ours}S. Andr\'e, V. Brosco, A. Shnirman, and G. Sch\"on, Phys. Rev. A. {\bf 79}, 053848 (2009)
\bibitem{mandel} P. Mandel, Phys. Rev. A {\bf 21},   2020 (1980).
\bibitem{PhysRev.188.1969}
B.~R. Mollow, Phys. Rev. {\bf 188}, 1969 (1969).
\bibitem{Bloch_Derivation}
F.~Bloch, Phys. Rev. {\bf 105},  1206 (1957).
\bibitem{Redfield_Derivation}
A.~G. Redfield, IBM J. Res. Dev. {\bf 1}, 19 (1957).
\bibitem{Zakrzewski}
J.~Zakrzewski, M.~Lewenstein, and T.~W. Mossberg,
Phys. Rev. A {\bf 44}, 7717 (1991).
\bibitem{Saclay_Karlsruhe}
G.~Ithier  \emph{et al.},  Phys. Rev. B {\bf 72}, 134519 (2005).
\bibitem{hauss08}J. Hauss \emph{et al.}, Phys. Rev. Lett. {\bf 100},  037003
(2008).

\bibitem{tavis} M. Tavis and F.W. Cummings, Phys. Rev. {\bf 170}, 379 (1968).
\bibitem{Gardiner} C. W. Gardiner and  P. Zoller, \emph{Quantum Noise}, Springer, Berlin, 2004.
\bibitem{cohen} C. Cohen-Tannoudji, J. Dupont-Roc, and G. Grynberg,
        \emph{Atom-Photon Interactions}, (Wiley, New York, 1992).

\bibitem{Ashhab} S. Ashhab \emph{et al.}, New J. Phys. {\bf 11}, 023030 (2009)
\bibitem{mu} Y. Mu and C.M. Savage, Phys. Rev. A {\bf 46},   5944 (1992).
\bibitem{briegel}C. Ginzel \emph{et al.}, Phys. Rev. A {\bf 48},   732
(1993).
\bibitem{rodrigues07-1}D.A. Rodrigues, J. Imbers, and  A.D. Armour, Phys.
Rev. Lett. {\bf 98},  067204 (2007).
\bibitem{falci} G. Falci \emph{et al.}, Phys. Rev. Lett. {\bf 94}, 
167002 (2005).
\bibitem{wang} Z. C. Wang and H. Haken,  Z. Phys. B {\bf 55}, 361 
(1984).



\end{thebibliography}
\end{document}